\documentclass{emulateapj}
\usepackage{graphicx}
\usepackage{natbib}
\usepackage{amsmath}
\slugcomment{ }
\begin{document}

\title{Confirmation of Hostless Type Ia Supernovae Using Hubble Space Telescope Imaging}

\author{
M.~L. Graham\altaffilmark{1},
D.~J. Sand\altaffilmark{2},
D. Zaritsky\altaffilmark{3},
C.~J. Pritchet\altaffilmark{4}
}

\altaffiltext{1}{Department of Astronomy, University of California, Berkeley, CA 94720-3411 USA}
\altaffiltext{2}{Physics Department, Texas Tech University, Lubbock, TX 79409}
\altaffiltext{3}{Steward Observatory, University of Arizona, Tucson AZ 85721}
\altaffiltext{4}{Department of Physics and Astronomy, University of
Victoria, PO Box 3055, STN CSC, Victoria BC V8W 3P6, Canada}

\begin{abstract}
We present deep Hubble Space Telescope imaging at the locations of four, potentially hostless, long-faded Type Ia supernovae (SNe\,Ia) in low-redshift, rich galaxy clusters that were identified in the Multi-Epoch Nearby Cluster Survey. Assuming a steep faint-end slope for the galaxy cluster luminosity function ($\alpha_d=-1.5$), our data includes all but $\lesssim0.2\%$ percent of the stellar mass in cluster galaxies ($\lesssim0.005\%$ with $\alpha_d=-1.0$), a factor of 10 better than our ground-based imaging. Two of the four SNe\,Ia still have no possible host galaxy associated with them ($M_R>-9.2$), confirming that their progenitors belong to the intracluster stellar population. The third SNe\,Ia appears near a faint disk galaxy ($M_V=-12.2$) which has a relatively high probability of being a chance alignment. A faint, red, point source coincident with the fourth SN\,Ia's explosion position ($M_V=-8.4$) may be either a globular cluster (GC) or faint dwarf galaxy. We estimate the local surface densities of GCs and dwarfs to show that a GC is more likely, due to the proximity of an elliptical galaxy, but neither can be ruled out. This faint host implies that the SN\,Ia rate in dwarfs or GCs may be enhanced, but remains within previous observational constraints. We demonstrate that our results do not preclude the use of SNe\,Ia as bright tracers of intracluster light at higher redshifts, but that it will be necessary to first refine the constraints on their rate in dwarfs and GCs with deep imaging for a larger sample of low-redshift, apparently hostless SNe\,Ia.
\end{abstract}


\section{Introduction}\label{s:intro}

Supernovae of Type Ia (SNe\,Ia) are the thermonuclear explosions of carbon-oxygen white dwarf (CO WD) stars, commonly used as cosmological standard candles although their progenitor scenario is not yet well understood (e.g., Howell 2011). Most likely, the WD is in a binary system with either another WD or a red giant or main sequence star, and the explosion occurs after merger with, or sufficient mass accretion from, the companion. The explosion rate of SNe\,Ia is correlated with galaxy mass and star formation rate, and most of the discovered SNe\,Ia reside in large galaxies -- but since SNe\,Ia are very bright, they are also used as ``cosmic lighthouses" for faint or diffuse astrophysical structures that are difficult to assess directly. The purpose of this work is threefold: (1) to address the utility of SNe\,Ia as a bright tracer of baryons in rich galaxy clusters, (2) to confirm that SNe\,Ia progenitors include truly old ($>2$ Gyr) progenitor systems, and (3) to investigate constraints on the SN\,Ia rate in faint hosts such as dwarf galaxies and globular clusters (GC). We motivate each of these science goals in turn.

\subsection{Baryon Accounting}\label{ssec:intro.baryons}

Understanding the growth of structure in the universe requires a full accounting of baryonic mass, and this must include the population of intracluster (IC) stars that were stripped from their host galaxy and reside in the cluster potential. Direct measurements of this low surface brightness component are possible (e.g., Gonzalez et al. 2005; Zibetti et al. 2005; Montes \& Trujillo 2014; DeMaio et al. 2015), but are difficult beyond the local universe due to cosmological surface brightness dimming. Indirect measurements of the fraction of intracluster light ($f_{\rm ICL}$) can be made using bright tracers of the underlying stellar population such as planetary nebulae and novae, which has been done for the nearby Virgo and Fornax clusters respectively (Feldmeier et al. 2004; Neill et al. 2005). At higher redshifts a brighter tracer is required, and $f_{\rm ICL}$ can instead be calculated by comparing the number of SNe\,Ia hosted by cluster galaxies to the number that appear hostless and belong to the IC stellar population.

This was first done with the Wise Observatory Optical Transients Search (WOOTS) by Gal-Yam et al. (2003), who found that 2 of their 7 SNe\,Ia in rich, low-redshift galaxy clusters appeared to be hostless, which implied that $f_{\rm ICL}\approx20\%$. A similar measurement was made with the Sloan Digital Sky Survey (SDSS) by McGee \& Balogh (2010), who found that 19 of the 59 SNe\,Ia in low-redshift galaxy groups appeared to be hostless, which implied that $f_{\rm ICL}\approx50\%$ for smaller-scale structures (galaxy groups have a total mass $\sim10\%$ that of rich clusters). Most recently, the Multi-Epoch Nearby Cluster Survey (MENeaCS) from the Canada France Hawaii Telescope (CFHT) identified four apparently hostless SNe\,Ia in low-redshift massive galaxy clusters (Sand et al. 2011), which implied that $f_{ICL} = 0.16_{-0.09}^{+0.13}$. 

This usefulness of SNe\,Ia as tracers of the ICL depends on their being {\it truly} hostless. Stacked imaging from the past surveys left $0.03-0.3\%$ (WOOTS), $3\%$ (SDSS), and $0.05-0.1\%$ (MENeaCS) of the stellar mass in cluster galaxies below the detection thresholds. These values are dependent on the logarithmic faint-end slope of the luminosity function, $\alpha_d$ (i.e. the Schechter function). The above results are based on $\alpha_d=-1.0$, which is true for field galaxies (Blanton et al. 2003), but $\alpha_d$ may be higher in rich galaxy clusters (e.g., Milne et al. 2007). Adopting $\alpha_d=-1.5$ as an upper limit, the MENeaCS survey estimated that $\lesssim2\%$ of the stellar mass in dwarf cluster galaxies below detection thresholds. If the SN\,Ia occurrence rate per unit mass is equivalent in all cluster stellar populations -- detected galaxies ($\sim82\%$), undetected dwarf galaxies ($\lesssim2\%$), and intracluster stars ($\sim16\%$) -- then for MENeaCS we expect to find that $\lesssim2\%$ of all the discovered cluster SNe\,Ia (23) are hosted by faint dwarf galaxies ($\sim0.5$ SNe\,Ia). In this work we use deep $HST$ imaging at the locations of the 4 MENeaCS IC SNe\,Ia to further lower the fraction of undetected mass in faint galaxies to $0.2\%$ ($0.003$--$0.007\%$ with $\alpha_d=-1.0$), analyze previously undetected objects in the vicinity of 2 SNe\,Ia, and discuss how our results affect the ability of IC SNe\,Ia to measure $f_{\rm ICL}$ at higher redshifts.

\subsection{SN\,Ia Delay Times}

One path towards understanding the progenitor scenario of SNe\,Ia is to constrain the range of possible ages via the SN\,Ia delay time distribution (DTD; the time between star formation and explosion). While Type II and Ib/c supernovae -- the core collapse of massive stars -- are associated with young stellar populations, SNe\,Ia occur with an explosion rate proportional to both galaxy stellar mass and star formation rate (e.g., Mannucci et al. 2005; Scannapieco \& Bildsten 2005; Sullivan et al. 2006). This indicates that SNe\,Ia are associated with both old and young stellar populations. The current best measurements of the SN\,Ia DTD indicate that some SN\,Ia progenitors are quite old, $>$2 Gyr (e.g. Maoz, Mannucci \& Nelemans 2014), which implies long-lived progenitors and/or that the timescales for mass transfer or merger are long. 

The conclusion that at least some SN\,Ia require progenitors that are $>2$ Gyr old relies on the SN\,Ia rate in rich cluster galaxies, where the majority of the stellar population is old (e.g. Sand et al. 2012). However, there is evidence that low levels of star formation are present in elliptical galaxies (e.g., Yi et al. 2005; Suh et al. 2011; Graham et al. 2012). Could it be that all SN\,Ia are actually from younger stellar populations? Probably not -- for example, Graham et al. (2012) show that the SN\,Ia DTD result is robust to this small amount of young stars. Even so, direct confirmation of a SN\,Ia progenitor in a stellar population of purely old stars would strengthen and support the late-time DTD constraints on the progenitor scenario. 

A suitable environment for this test is the population of intracluster (IC) stars in rich galaxy clusters: the colors and luminosities of IC red giant stars in Virgo show that the IC stellar population is comprised of stars $\geq$ 2 Gyr old (Durrell et al. 2002), and theoretical models also suggest the ICL is comprised of old stars (e.g. Sommer-Larsen et al. 2005; Purcell et al. 2007). All of the apparently hostless cluster SNe discovered to date are of Type Ia; a lack of intracluster core collapse supernovae lends credence to the idea that the ICL is composed of an older stellar population. In this work, we use deep $HST$ imaging to show that at least two of the apparently hostless SNe\,Ia discovered by MENeaCS truly belong to the IC stellar population of old stars.

\subsection{SN\,Ia Rates in Faint Hosts}

Over the past ten years, an increasing number of SN surveys have employed an unbiased, wide-field search strategy instead of targeting massive galaxies. This has lead to the discovery that some types of luminous SNe have a higher explosion rate in dwarf hosts (e.g. Neill et al. 2011; Lunnan et al. 2014), attributed to high star formation rates providing more progenitor stars and/or lower metallicity leading to more luminous SNe, or perhaps an elevated rate of binary star formation. As most SNe\,Ia occur in massive galaxies, it is difficult to assess whether they might also have an enhanced rate in dwarf galaxies because of the large statistical uncertainty on the rate due to the relatively low number of SNe\,Ia in low-mass hosts (e.g. Quimby et al. 2012). SN surveys of rich galaxy clusters are an efficient way to search many low-mass galaxies at once, both due to a higher sky density of galaxies and the (putative) upturn in the faint-end slope of the cluster luminosity function (e.g., Milne et al. 2007; Yamanoi et al. 2007). In this work we find that one IC SN\,Ia may be hosted by a dwarf cluster galaxy, and discuss the implication of this for SN\,Ia rates in faint hosts.

GCs are another potential very faint host for IC SNe\,Ia. No SN has ever been confirmed to be hosted by a GC, although they are theoretically predicted to have a SN\,Ia occurrence rate $1-10\times$ that of elliptical galaxies due to dynamical interactions that lead to more white dwarfs in binary systems in GCs (e.g., Ivanova et al. 2006; Shara \& Hurley 2006; Pfahl et al. 2009). Non-detections of GCs at the sites of $\sim45$ low-redshift SNe\,Ia in archival $HST$ images have constrained the potential rate enhancement to $\lesssim42\times$ the rate in elliptical galaxies (Voss \& Nelemans 2012; Washabaugh \& Bregman 2013). In this work we find that one IC SN\,Ia may be hosted by a GC, and discuss the implication of this for SN\,Ia rates in GCs. Since GCs are also comprised mainly of old (5--13) Gyr stellar populations, confirming a GC-hosted SNe\,Ia would also meet our science goal of confirming SNe\,Ia with long ($>2$ Gyr) delay times. 

\subsection{Paper Overview}

In this work, we use the Hubble Space Telescope ({\it HST}) Advanced Camera for Surveys (ACS) to obtain deep images in filters F606W and F814W at the locations of 4 apparently hostless SNe\,Ia from MENeaCS (Sand et al. 2011). These are the deepest images, and the largest single survey sample, of IC SNe\,Ia locations in rich clusters ever obtained and analyzed in this way. In Section \ref{sec:obs} we present our original CFHT and new \textit{HST} observations and discuss our image processing and photometric calibrations. In Section \ref{sec:ana} we analyze our deep stacks of {\it HST} ACS imaging: we characterize the faint sources in the vicinity of the SNe\,Ia, derive our point-source limiting magnitudes, and determine the amount of cluster stellar luminosity remaining below our detection thresholds. In Section \ref{sec:disc} we discuss the implications of these results with respect to SN\,Ia progenitor ages, the rates of SNe\,Ia in faint cluster objects, and the future use of SNe\,Ia as tracers of the ICL. We conclude in Section \ref{sec:conc}. All dates are given in UT and a standard flat cosmology of $\Omega_M=0.3$, $\Omega_{\Lambda}=0.7$ is assumed throughout.

\section{Observations}\label{sec:obs}

Here we describe our past observations with the MegaCam imager (Boulade et al. 2003) at the Canada-France-Hawaii Telescope (CFHT), and our new deep imaging taken 4--5 years later with the Hubble Space Telescope \textit{HST} Advanced Camera for Surveys (ACS; Holland et al. 1998).

\subsection{CFHT MENeaCS}

The Multi-Epoch Nearby Cluster Survey (MENeaCS) monitored 57 low-redshift ($0.05<z<0.15$) massive galaxy clusters from 2008-2010 with the MegaCam instrument at the Canada-France-Hawaii Telescope. A total of 23 cluster SNe\,Ia (Sand et al. 2012) and 7 cluster SNe\,II (Graham et al. 2012) were discovered. The survey strategy, spectroscopic follow-up, detection efficiencies, and the derivation of SN rates from MENeaCS are presented in Sand et al. (2012), and the four intracluster SNe\,Ia discovered by MENeaCS are presented in Sand et al. (2011). For the 4 IC SNe\,Ia, in Table \ref{tab:icsneia} we list their MENeaCS identifier, coordinates, spectroscopic redshift, and type as determined by the Supernova Identification (SNID) software package (Blondin \& Tonry 2007). As described in Sand et al. (2011), these SNe\,Ia are all within 1 Mpc of the brightest cluster galaxy and within 3000 $\rm km\ s^{-1}$ of the cluster redshift, thereby confirming they occurred in the cluster. All four were more than 5 effective radii (i.e., $>5\times$ the half-light radius) from the nearest potential host galaxy in the CFHT images. Deep stacks were made from SN-free CFHT survey images, and implanted simulated point sources were used to constrain the limiting magnitude of any possible host galaxy. These limits are listed for each IC\,SN\,Ia in Table \ref{tab:icsneia}, along with an upper limit on the fraction of light in low-mass cluster galaxies below our detection limit ( $f(<L_{min})$). Assuming a faint-end slope for the cluster luminosity function of $\alpha_d=-1.5$, the fraction of stellar mass below the detection thresholds of our CFHT deep stacks is $\lesssim$2\% (Sand et al. 2011).

\begin{table*}
\begin{center}
\caption{CFHT Data for MENeaCS Intracluster Supernovae\tablenotemark{a} \label{tab:icsneia}}
\begin{tabular}{lcccccc}
\hline
MENeaCS        & SN Coordinates    & BCG Offset  & Redshift &  Spectral  & CFHT Detection Limit   & $f(<L_{min})$    \\
Identifier            & RA, Dec               & (kpc)             & (SNID)    & Type        & $M_g$, $M_r$    & $\alpha_d = -1.5$  \\
\hline
Abell1650\_9\_13\_0  & 12:59:01.33, $-$01:45:51.68  & 468 & 0.0836 &  Ia-norm  & $-12.47$, $-13.04$  & 0.0172  \\
Abell2495\_5\_13\_0  & 22:50:26.33, +10:54:41.70 & 148 & 0.0796 &  Ia-norm  & $-11.72$, $-12.37$  & 0.0127  \\
Abell399\_3\_14\_0    & 02:57:26.41, +12:58:07.63 & 616 & 0.0613 & Ia-norm  & $-12.54$, $-12.56$  & 0.0138  \\
Abell85\_6\_08\_0      & 00:42:02.39, $-$09:26:58.00  & 595 & 0.0617 &  Ia-91bg  & $-11.15$, $-11.68$  & 0.0091  \\
\hline 
\end{tabular}
\tablenotetext{1}{From Table 1 in Sand et al. (2011).}
\end{center}
\end{table*} 

\subsection{HST ACS Imaging}

To assess whether these four apparently hostless SNe\,Ia truly belonged to the intracluster stellar population we used the {\it HST} ACS to obtain deep images at their locations. We waited until $>$3 years after peak brightness to obtain these images to avoid contamination from the SN itself (this is discussed in Section \ref{ssec:snlim}). To either rule out or classify a faint object at the SN position we required an imager that offers both sensitivity and high resolution. Although the Wide Field Camera 3 (WFC3) has slightly better resolution, we chose the ACS for its higher throughput and increased sensitivity beyond 4000 \AA\ because most objects in galaxy clusters are red. At our target position on the CCD we selected aperture ``WFC" because amplifier ``B" on WFC1 has the lowest read noise. We restricted our telescope orientations to avoid possible stellar diffraction spikes coming near our target coordinates. Because a photometric color is necessary to classify a detected object we used two wide filters: F606W ($V$-band) and F814W ($I$-band). We divided our integration times into multiple exposures to remove cosmic rays and used small dithers to mitigate the effect of hot pixels, and were able to obtain all observations of a given cluster in a single visit. We provide a summary of our \textit{HST} observations in Table \ref{tab:hstobs}.

\begin{table*}
\begin{center}
\caption{HST+ACS Imaging Data for MENeaCS Intracluster Supernovae \label{tab:hstobs}}
\begin{tabular}{lcccccccc}
\hline
MENeaCS  & Orbits & Observation &  \multicolumn{2}{c}{Exposure Time (s)} & \multicolumn{3}{c}{HST Detection Limits (mag)}  & $f(<L_{min})$ \\
Identifier &  &  Date (UT) &  F606W & F814W   &  F606W & F814W & $M_{\rm R}$ & $\alpha_d = -1.5$                         \\
\hline
Abell1650\_9\_13\_0  & 5 & 2013-01-25 &  3870 & 6120  & 28.58 & 28.88  &  $-9.7$  & 0.0029 \\
Abell2495\_5\_13\_0  & 4 & 2013-10-04 &  2984 & 5970  & 28.64 & 28.90  & $-9.8$  & 0.0026 \\
Abell399\_3\_14\_0    & 2 & 2013-12-04 & 1800  & 1980  & 28.30  & 28.45 &  $-10.1$ & 0.0035 \\
Abell85\_6\_08\_0      & 1 & 2013-09-18 & 1800  &            &  28.40 &             & $-9.2$   & 0.0020 \\
                                       & 1 & 2014-09-05 &             & 1800  &            & 28.49 &         &  \\
\hline 
\end{tabular}
\end{center}
\end{table*} 

To make the deepest possible stacks of our \textit{HST}+ACS data, we use the {\it Astrodrizzle} software to median-combine FLC images provided by that STScI pipeline (Gonzaga et al. 2012). FLC files come fully reduced, drizzled, and corrected for charge transfer efficiency by the STScI pipeline. We use the WCS astrometry during image combination. We do this for the F606W and F814W images separately to make a deep stack for each filter that is free of cosmic rays, and then also create a single sum-combine image for the deepest possible stack.

\subsection{{\it HST} ACS Photometric Calibrations}\label{ssec:phot}

To obtain the apparent magnitude of a source, $m_f$, where $f$ represents filter F606W or F814W, we start with the raw magnitude, $m_{f,{\rm raw}}$, for which we use MAG\_AUTO from Source Extractor (Bertin \& Arnouts 1996). We add a small PSF correction for point sources, derived from the simulated sources used for the limiting magnitudes ($\sim -0.1$, see Section \ref{ssec:lim}). This is not usually necessary with MAG\_AUTO but we find it is required for the non-Gaussian PSF of the $HST$ ACS (the Tiny Tim PSF; Krist et al. 2011). We also apply the zeropoint, $z_f= -2.5 \log{P_f} -21.1$, where $P_f$ is the PHOTFLAM header keyword, representing the flux of a source with constant $\rm F_{\lambda}$ which produces a count rate of 1 electron per second ($P_f$ has units of $\rm erg\ cm^{-2}\ s^{-1}\ \AA^{-1}$). For our images, $z_{\rm F606W} = 26.66$ and $z_{\rm F814W} = 26.78$ mag. To transform from the natural system and obtain the apparent magnitude, $m_F$, where $F$ is the Johnson V or Cousins I filters, we use this equation from Sirianni et al. (2005; see their Equation 12):

\begin{equation}
m_F  = m_{f,{\it raw}} + C_{0,F} + (C_{1,F} \times {\rm TCOL}) + (C_{2,F} \times {\rm TCOL}^2)
\end{equation}

\noindent
where TCOL is the color in the targeted system; in our case ${\rm TCOL} = m_{\rm V} - m_{\rm I}$. We use the synthetic coefficient values from Sirianni et al. (2005; see their Table 22), as listed in our Table \ref{tab:trans}.

\begin{table}
\begin{center}
\caption{Photometric Transformations}
\label{tab:trans}
\begin{tabular}{llcccc}
\hline
Natural   & Target        &  $C_0$  & $C_1$ & $C_2$ & TCOL \\
Filter       & Filter          & & & & $m_{\rm V}-m_{\rm I}$ \\
\hline
F606W   & Johnson V & 26.394 & $0.153$ & $0.096$ & $<0.4$  \\
F606W   & Johnson V & 26.331 & $0.340$ & $-0.038$ & $>0.4$ \\
F814W   & Cousins I   & 25.489 & $0.041$ & $-0.093$ & $<0.1$ \\
F814W   & Cousins I   & 25.496 & $-0.014$ & $0.015$ & $>0.1$ \\
\hline  
\end{tabular}
\end{center}
\end{table} 

\section{Analysis}\label{sec:ana}

In this section we analyze our deep {\it HST} ACS images at the locations of our 4 intracluster SNe\,Ia from MENeaCS. We determine our limiting point-source magnitudes in Section \ref{ssec:lim}, and derive the fraction of cluster luminosity remaining below our detection thresholds in Section \ref{ssec:fL}. We present and characterize any objects found near the SN\,Ia coordinates in each cluster in Section \ref{ssec:coreg}, quantify the likelihood that detected objects are the result of a chance alignment in Section \ref{ssec:chance}, and rule out the possibility of observing the evolved companion star or lingering SN\,Ia emission in Section \ref{ssec:snlim}.

\subsection{Point Source Limiting Magnitudes}\label{ssec:lim}

\begin{figure}
\begin{centering}
\includegraphics[trim=0.8cm 0.0cm 0.4cm 0.0cm, clip, width=8.5cm]{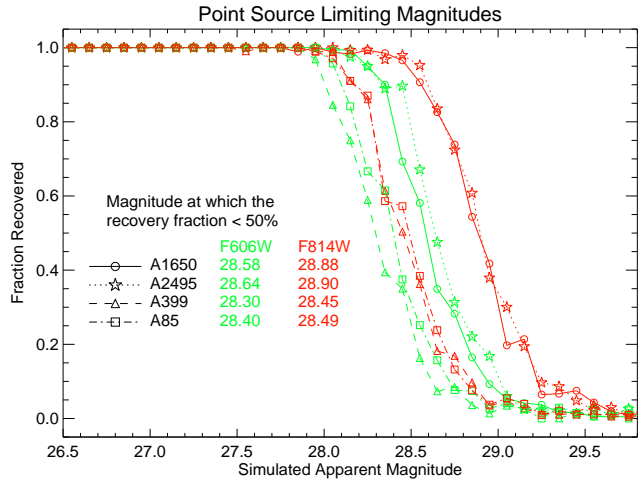}
\caption{Detection efficiencies for simulated point sources in our \textit{HST} ACS images, as described in the text (\S~\ref{ssec:lim}). Filter and cluster are represented by color and symbol respectively, as shown in the plot legend. The magnitude at which our detection efficiency falls to 50\% is considered our limiting magnitude. \label{fig:de}}
\end{centering}
\end{figure}

In order to determine the limiting magnitude of our images, we plant 5000 fake stars in each of our stacked F606W and F814W images. These fake stars have apparent magnitudes $26.0 < m < 30.0$, with more stars at fainter magnitudes, and an appropriate PSF from the Tiny Tim model PSF generator (Krist et al. 2011). We ensure that simulated stars are only planted in regions of low surface brightness in order to mimic the locations of our intracluster SNe\,Ia. We run Source Extractor on the images with the simulated population of point-sources; the relevant detection parameters are given in Table \ref{tab:sexpars}. We visually verify that these parameters are returning all and only real sources in the images. In Figure \ref{fig:de} we plot our detection efficiency (i.e. the fraction of objects recovered) as a function of apparent magnitude for the F606W and F814W images of each cluster. The ``limiting magnitude" is defined as the magnitude at which the detection efficiency drops to 50\%, and is given for each cluster and filter in the plot legend of Figure \ref{fig:de}.

In Section \ref{ssec:coreg} below, we visually identify objects below the official limiting point source magnitude at the locations of our SNe\,Ia in Abell 2495 and Abell 399. We find that these objects are only detected by Source Extractor if the threshold is lowered to $1\sigma$, which also detects many peaks in the background noise and produces a source catalog with large uncertainties in their apparent magnitude. This is why our official limiting magnitude -- which needs to be robust because we use it to determine the fraction of cluster stellar luminosity below our detection thresholds in Section \ref{ssec:fL} -- is slightly brighter than some of the sources discussed in Section \ref{ssec:coreg}. The caveat here is that brighter, but more extended, objects may fall below our detection threshold also -- but most of the faint {\it cluster} objects will be point-like (dwarf galaxies and GCs).

\begin{table}
\begin{center}
\caption{Source Extractor Parameters \\ for the Detection Efficiency}
\label{tab:sexpars}
\begin{tabular}{lc}
\hline
Parameter                       & Value   \\
\hline
DETECT\_MINAREA      & 3          \\
DETECT\_THRESH       & 2          \\
SEEING\_FWHM           &  0.08       \\
BACK\_SIZE	               &  256	   \\
BACK\_FILTERSIZE       &  5           \\
BACK\_TYPE                 &  AUTO    \\
BACKPHOTO\_TYPE    &  LOCAL   \\	   
\hline  
\end{tabular}
\end{center}
\end{table} 

\subsection{Fraction of Undetected Cluster Light}\label{ssec:fL}

To determine the fraction of cluster light remaining below our point-source limiting magnitudes, we follow a similar method to that presented by Sand et al. (2011) for our CFHT deep-stack images, the result of which is reproduced in the last column of Table \ref{tab:icsneia}.

The absolute $R$-band luminosity function for the nearby Virgo cluster is modeled by Trentham \& Tully (2002) with two components, a Gaussian for the bright end and a Schechter function (see their Equation 2):
\begin{multline}
N(M) = N_g e^{-(M-M_g)^2/(2\sigma_g^2)} \\ + N_d (10^{[-0.4(M-M_d)]})^{\alpha_d+1} e^{-10^{[-0.4(M-M_d)]}},
\end{multline}
\noindent
where $N(M)$ is the number density of galaxies per square $\rm Mpc$ per magnitude bin, $M$ is the $R$-band absolute magnitude, and $N_g=17.6$, $M_g=-19.5$, $\sigma_g=1.6$, $N_d=3N_g$, $M_d=-18.0$, and $\alpha_d=-1.03$. However, the faint-end slope is known to steepen with redshift (e.g., Khochfar et al. 2007), and values down to $\alpha_d \approx-1.5$ have been measured for the Coma cluster (Milne et al. 2007) and most recently for Abell 85 ($\alpha_d\approx -1.6$; Agulli et al. 2014). In order to make a robust upper limit on the amount of cluster light below our detection efficiencies, we use $\alpha_d=-1.5$ from here on.

In Sand et al. (2012) the total $r^{\prime}$-band luminosity is calculated for all MENeaCS clusters. We convert this to $R$-band using the conversion factors from Blanton \& Roweis (2007), $R = r - 0.0576 -0.3718( (r-i) - 0.2589)$. Because most of the cluster light is from old stellar populations, we use the typical SDSS color of elliptical galaxies, $r-i\sim0.4$, from Eisenstein et al. (2001). We integrate the galaxy luminosity function down to the CFHT point-source limiting magnitudes listed in Table \ref{tab:icsneia} from Sand et al. (2011), and then normalize to the total $R$-band luminosity for each cluster. 

We convert the absolute $R$-band luminosity function for each cluster into apparent $V$- and $I$-band using the cluster's redshift, the elliptical galaxy template spectrum from Kinney et al. (1996), and the line-of-sight Galactic extinction for each cluster: $A_{\rm V,A1650}=0.047$, $A_{\rm V,A2495}=0.211$, $A_{\rm V,A399}=0.467$, $A_{\rm V,A85}=0.103$,  $A_{\rm I,A1650}=0.026$, $A_{\rm I,A2495}=0.116$, $A_{\rm I,A399}=0.256$, and $A_{\rm I,A85}=0.057$ (Schlafly \& Finkbeiner 2011; Schlegel, Finkbeiner \& Davis 1998). We convert the $V$ and $I$-band luminosity function into the {\it HST} ACS natural system filters using the transformations of Sirianni et al. (2005), as described in Section \ref{ssec:phot}. In Table \ref{tab:hstobs}, we report the point-source detection limit in absolute $M_{\rm R}$ magnitudes, and the fraction of the cluster's $R$-band stellar luminosity remaining below this limit. For all clusters we find that $<0.2\%$ of the stellar mass in cluster galaxies remains below our official point-source limiting magnitude. This result is discussed further in Section \ref{sec:disc}.

\subsection{Potential Hosts in the Deep \textit{HST}-ACS Images} \label{ssec:coreg}

\begin{figure*}
\begin{centering}
\includegraphics[trim=0.0cm 0.0cm 0.0cm 0.0cm, clip, width=18cm]{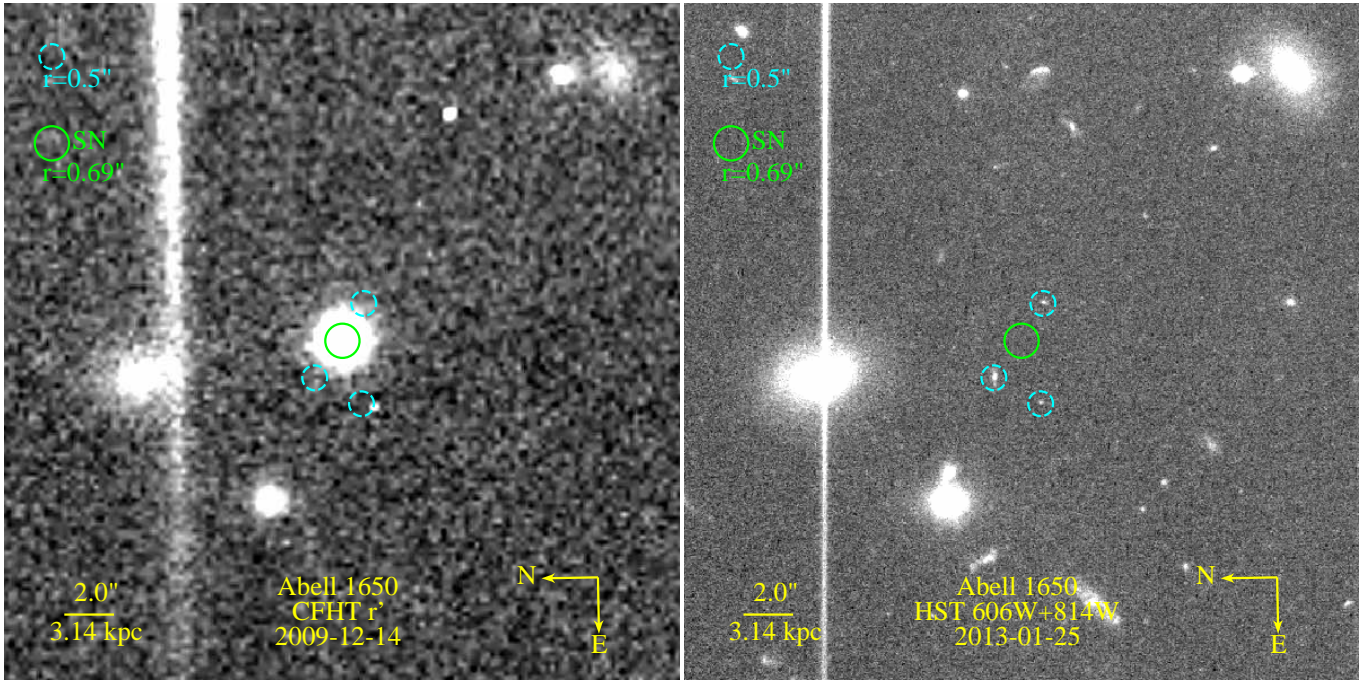}
\caption{Co-registered CFHT (left) and \textit{HST} (right) images for the SN\,Ia in Abell 1650, as described in the text of Section \ref{ssec:coreg}. \label{fig:coregA1650}}
\end{centering}
\end{figure*}

\begin{figure*}
\begin{centering}
\includegraphics[trim=0.0cm 0.0cm 0.0cm 0.0cm, clip, width=18cm]{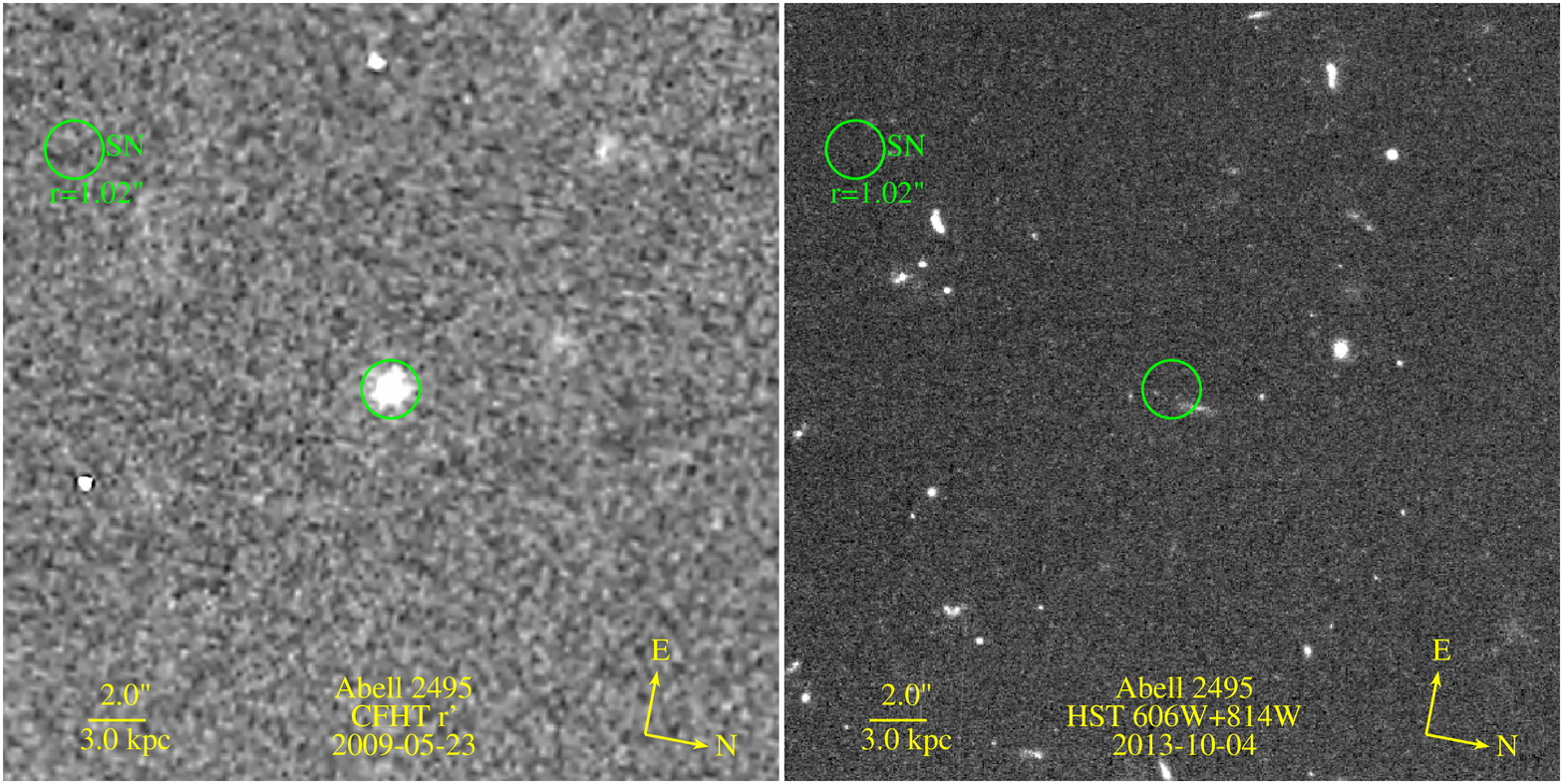}
\caption{Co-registered CFHT (left) and \textit{HST} (right) images for the SN\,Ia in Abell 2495, as described in the text of Section \ref{ssec:coreg}. \label{fig:coregA2495}}
\end{centering}
\end{figure*}

\begin{figure*}
\begin{centering}
\includegraphics[trim=0.0cm 0.0cm 0.0cm 0.0cm, clip, width=18cm]{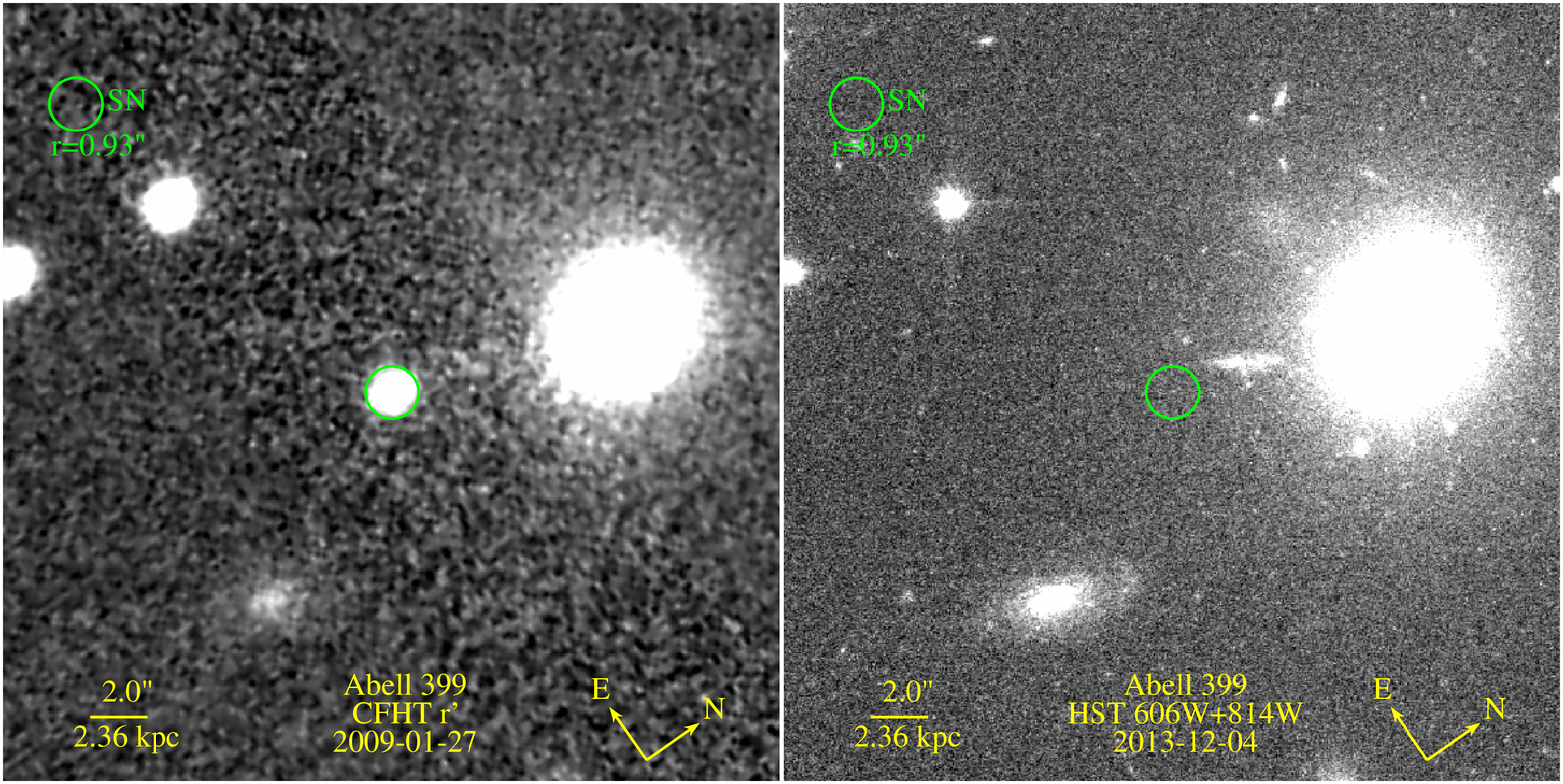}
\caption{Co-registered CFHT (left) and \textit{HST} (right) images for the SN\,Ia in Abell 399, as described in the text of Section \ref{ssec:coreg}. \label{fig:coregA399}}
\end{centering}
\end{figure*}

\begin{figure*}
\begin{centering}
\includegraphics[trim=0.0cm 0.0cm 0.0cm 0.0cm, clip, width=18cm]{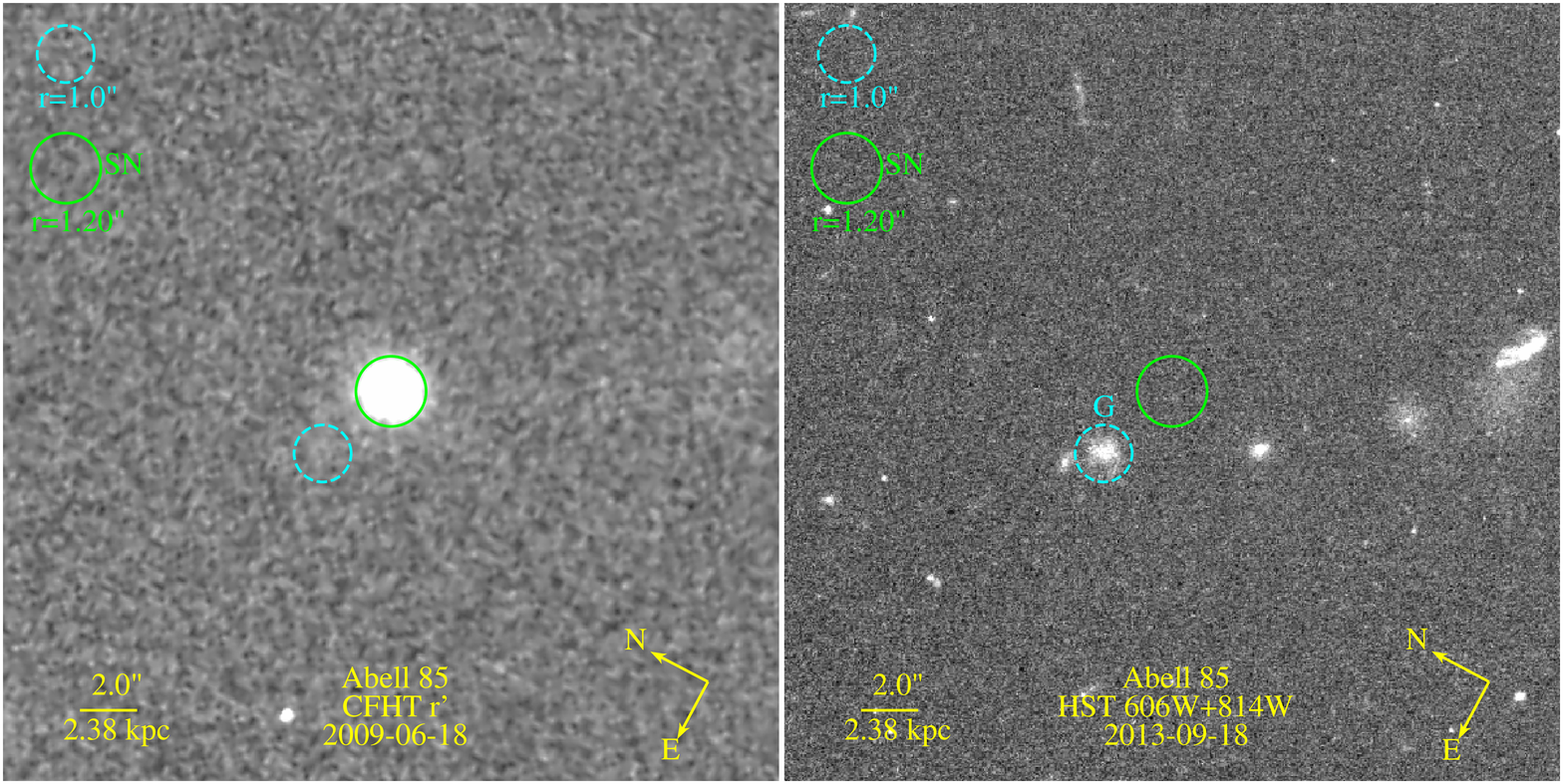}
\caption{Co-registered CFHT (left) and \textit{HST} (right) images for the SN\,Ia in Abell 85, as described in the text of Section \ref{ssec:coreg}. \label{fig:coregA85}}
\end{centering}
\end{figure*}

We use the IRAF tasks GEOMAP and GEOTRAN to co-register our CFHT images to the new, deeper \textit{HST}-ACS images. In Figures \ref{fig:coregA1650}--\ref{fig:coregA85} we show the results, side by side, for comparison. The CFHT images are comprised of two 120 second $r^{\prime}$ exposures, and contain the SN\,Ia -- we do not use our SN-free deep stacks here, because we need the SN's coordinates in the co-registered frames. The HST images are our deepest stacks, the sum-combined F606W and F814W filtered images. Green circles mark the position of the SN in each image, with a radius equal to 3$\times$ the positional uncertainty of the SN\,Ia. These positional uncertainties, listed in Table \ref{tab:snpos}, are a combination of Source Extractor's uncertainty in the PSF centroid for the SN in the co-registered CFHT image (using windowed output parameters), and the error in GEOMAP's transformation between images. The dashed cyan lines enclose nearby objects (sizes chosen to guide the eye). Image information such as the cluster name, UT date of acquisition, filters, scale bar, and compass are shown in yellow along the bottom. These images all have a 30\arcsec\ $\times$ 30\arcsec\ field of view.

\begin{table}
\begin{center}
\caption{Supernova Positional Uncertainty\tablenotemark{a}}
\label{tab:snpos}
\begin{tabular}{lccccccc}
\hline
SN\,Ia & \multicolumn{2}{c}{PSF Center\tablenotemark{b}} & \multicolumn{2}{c}{GEOMAP\tablenotemark{c}} & \multicolumn{2}{c}{Total\tablenotemark{d}} & Error \\
Cluster  & $\Delta x$ & $\Delta y$ & $\Delta x$ & $\Delta y$ & $\Delta x$ & $\Delta y$ & Radius\tablenotemark{e} \\
Abell  & (pix) & (pix) & (pix) & (pix) & (pix) & (pix) & (\arcsec) \\
\hline
1650  &  4.57  & 4.57 & 0.31 & 0.13  &  7.52  &  7.62  & 0.23\arcsec \\
2495  &  6.55  & 6.31 & 1.81 & 2.00  &  6.42  &  6.46  & 0.34\arcsec \\
399    &  6.15  & 6.00 & 0.63 & 1.56  &  7.76  &  7.82  & 0.31\arcsec \\
85      &  7.39  & 7.21 & 2.57 & 3.52  & 10.41 & 10.15 & 0.40\arcsec \\
\hline
\end{tabular}
\tablenotetext{1}{All pixels are \textit{HST} ACS pixels ($\rm 0.05\arcsec\ pixel^{-1}$).}
\tablenotetext{2}{Source Extractor's uncertainty on the SN's coordinates in the CFHT image post-transformation with GEOTRAN (i.e., the square root of the variance, the second moment of the barycenter).}
\tablenotetext{3}{GEOMAP's uncertainty in mapping the CFHT image to the \textit{HST} image (i.e., a systematic, the output $x_{rms}$ and $y_{rms}$ values).}
\tablenotetext{4}{Added in quadrature.}
\tablenotetext{5}{Average in $x$ and $y$, $\rm \times 0.05\arcsec\ pixel^{-1}$.}
\end{center}
\end{table} 

As in Sand et al. (2011), we use the dimensionless parameter $R$ to identify whether nearby objects could be considered as potential hosts of the SNe\,Ia. This parameter is defined in the Source Extractor manual as:

\begin{equation}\label{eq:r}
R^2 = C_{xx} x_r^2 + C_{yy} y_r^2 + C_{xy} x_r y_r
\end{equation}

\noindent
where $C_{xx}$, $C_{yy}$, and $C_{xy}$ are object ellipse parameters, $x_r = x_{SN} - x_{gal}$ and $y_r = y_{SN} - y_{gal}$, and $R\sim3$ describes the isophotal limit of the galaxy (see also Sullivan et al. 2006). A supernova is typically only classified as ``hostless" if $R>5$, but depending on the surface brightness profile of the galaxy, a significant amount of the stellar mass may reside beyond this radius (e.g., $\gtrsim10\%$ for the potential host of the SN in Abell 399, as determined in Sand et al. 2011). 

\subsubsection{Abell 1650}\label{sssec:a1650}

In Figure \ref{fig:coregA1650}, we see that the location of the SN\,Ia is truly devoid of objects. We ran Source Extractor with very relaxed parameters and still recovered no sources in this area. Of the three nearby objects enclosed by dashed circles in Figure \ref{fig:coregA1650}, the SN location is $R\gtrsim15$ away. Our intracluster SN\,Ia in Abell 1650 therefore appears to be truly hostless.

\subsubsection{Abell 2495}\label{sssec:a2495}

\begin{figure*}
\begin{centering}
\includegraphics[trim=0.0cm 0.0cm 0.0cm 0.0cm, clip, width=8cm]{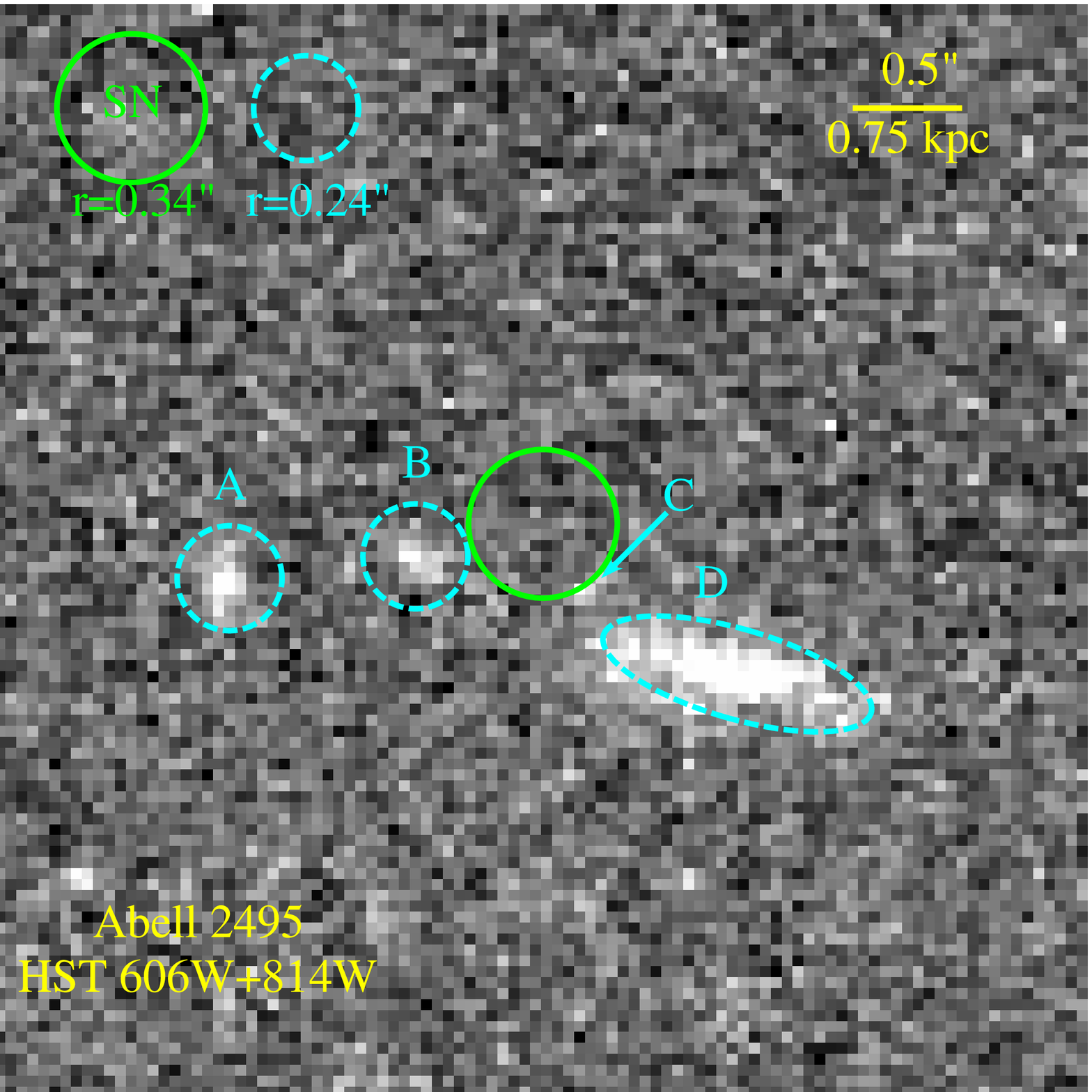}
\includegraphics[trim=0.3cm 0.0cm 0.3cm 0.0cm, clip, width=9.4cm]{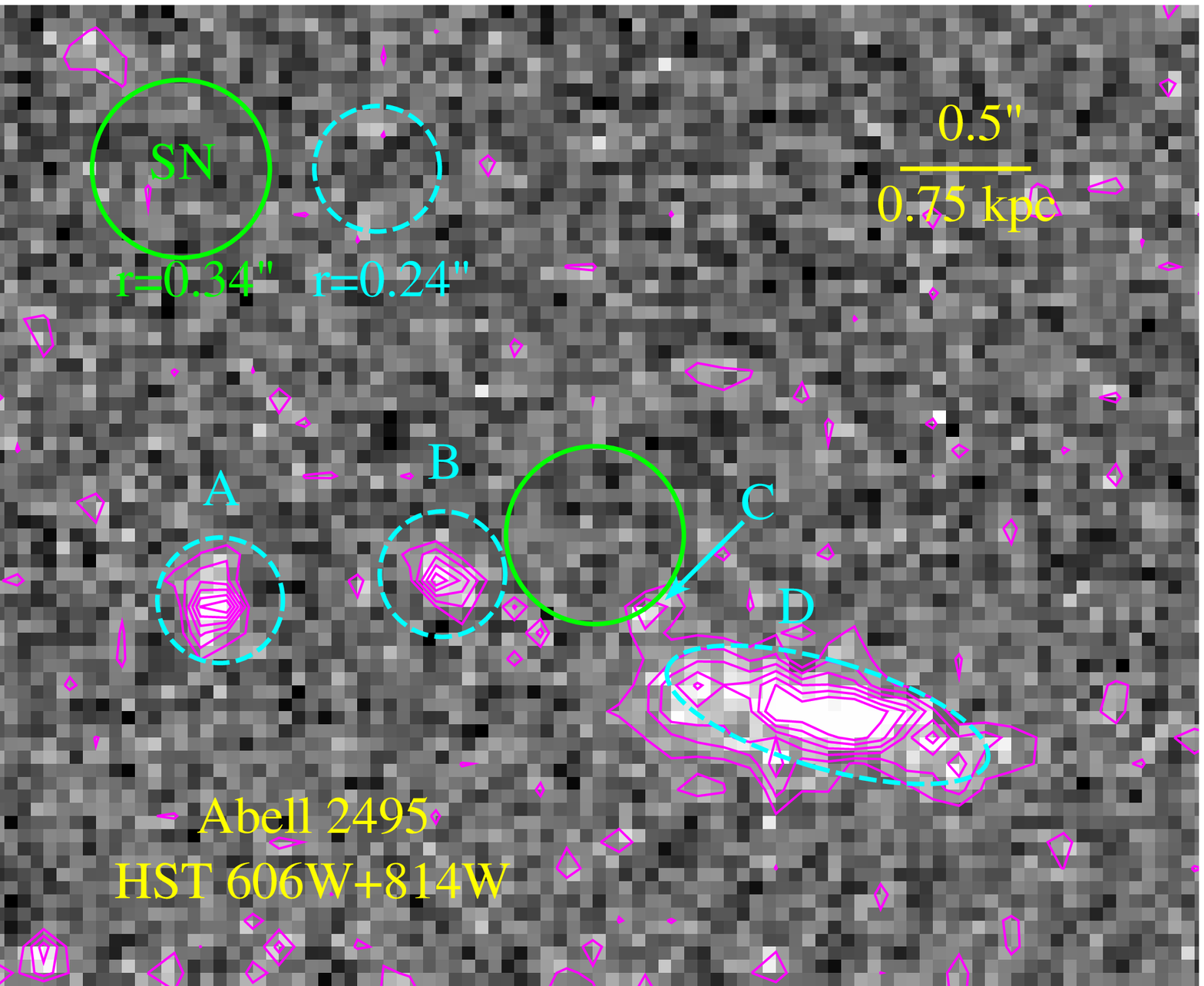}
\caption{The \textit{HST}+ACS image from the deepest stack of both filters F606W and F814W for the intracluster SN\,Ia in Abell 2495. The green circle marks the position of the SN\,Ia, with a radius equal to $1\times$ the positional uncertainty from Table \ref{tab:snpos}. In cyan we identify nearby objects A, B, C, and D (we cannot verify whether C is a part of D), discussed in the text. For objects A and B, the dashed cyan circle has $r=0.24\arcsec$, representing $3\times$ the PSF FWHM of ACS (0.08\arcsec). For the dashed ellipse around object D we use the parameters determined by Source Extractor for the F814W image: $3\times$ the semi-major and -minor axes, $A=4.4$ and $B=1.1$ pixels ($A=0.22\arcsec$\ and $B=0.053\arcsec$), and a position angle of $\theta=74$ degrees (in the x-y plane).  In the image at right, we include flux contours in magenta to highlight the connection between objects C and D and the lack of low-significance sources within the SN's positional error circle.  \label{fig:coregA2495zoom}}
\end{centering}
\end{figure*}

In Figure \ref{fig:coregA2495} we see several sources near the location of the SN\,Ia that were not apparent in the CFHT deep stacks presented in Sand et al. (2011), but none are within the positional uncertainty of the SN\,Ia. We ran Source Extractor with very relaxed parameters and still recovered no sources within the positional uncertainty. In Figure \ref{fig:coregA2495zoom} we show this region in detail, and identify nearby sources A, B, C, and D. Sources A and B are unlikely to be physically associated, as the SN\,Ia is $R\sim17$ away from them. We discuss objects C and D in turn.

\begin{figure}
\begin{centering}
\includegraphics[trim=0.9cm 0.1cm 0.4cm 0.0cm, clip, width=8.5cm]{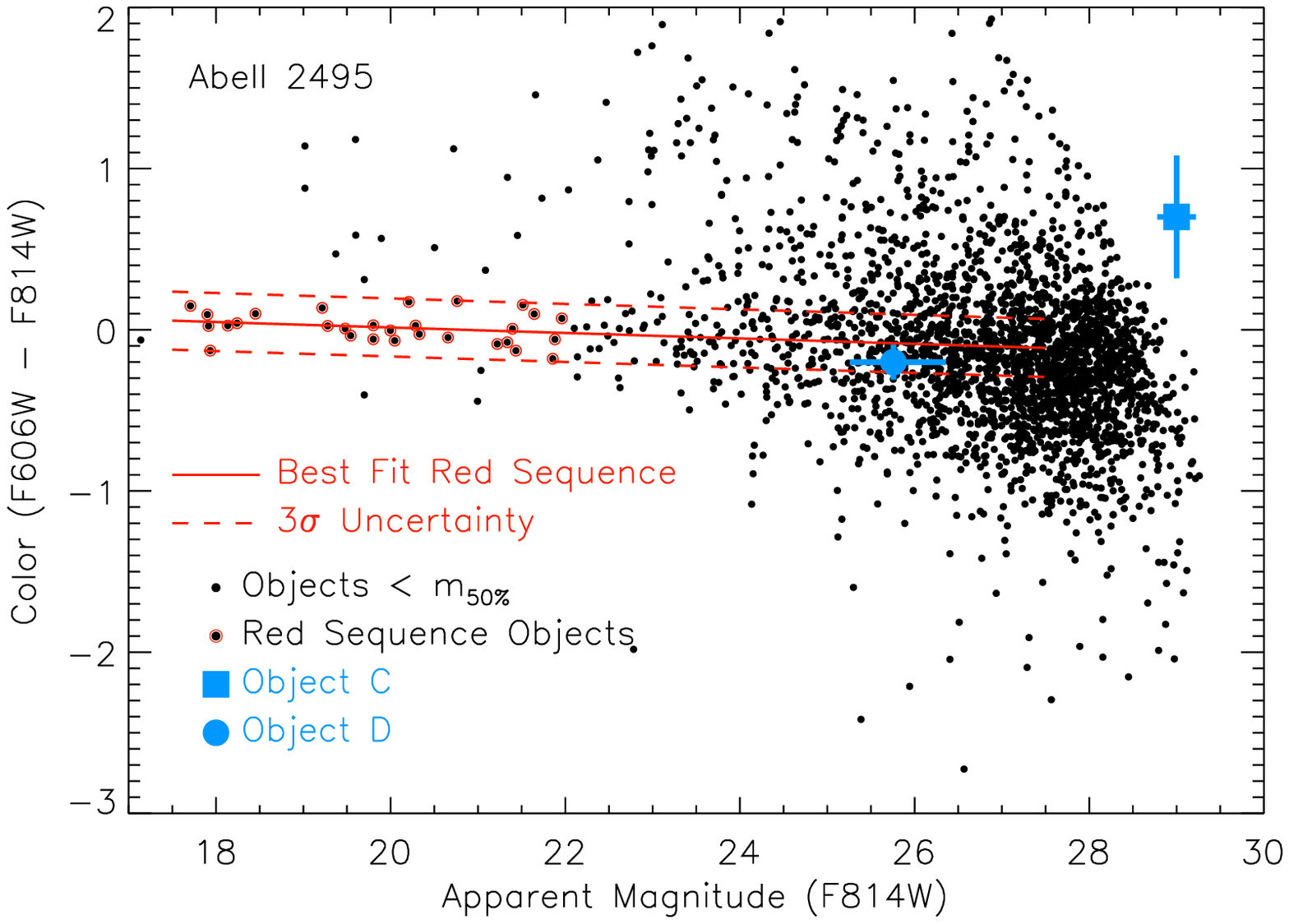}
\caption{Color-magnitude diagram in the natural \textit{HST} ACS filter system for Abell 2495. Black points represent all sources extracted with an apparent magnitude brighter than the 50\% detection efficiencies in both F606W and F814W, as discussed in \S~\ref{ssec:lim} and shown in Figure \ref{fig:de}. Black points surrounded by red indicate the galaxies used in the linear fit to the red sequence (red lines). The blue square and circle represent objects C and D respectively (see Figure \ref{fig:coregA2495zoom}).  \label{fig:A2495cmd}}
\end{centering}
\end{figure}

Object C is likely a part of object D, which is clumpy and extended, as shown by the contour plot in the right-hand image of Figure \ref{fig:coregA2495zoom}. However, object C is identified by Source Extractor as an independent source at the $1\sigma$ level in the F814W image, with $m_{\rm F814W} = 29.0 \pm 0.2$ mag. In the F606W image it is not officially detected by Source Extractor, but it is just visible to the eye, and with aperture photometry we estimate it to be $m_{\rm F606W} = 29.8 \pm 0.2$ mag. In both filters, object C falls below our 50\% detection efficiency for Abell 2495 (see Figure \ref{fig:de} in Section \ref{ssec:lim}). The SN's location is $R\gtrsim5$ away from object C, which argues against a physical association. Object C is redder than the red sequence of Abell 2495, as shown in Figure \ref{fig:A2495cmd}, and so it is unlikely to be a cluster dwarf galaxy. The scenarios in which object C is lingering emission from the SN\,Ia or a chance alignment are discussed in Sections \ref{ssec:chance} and \ref{ssec:snlim}. Ultimately we find it unlikely that object C is the host galaxy.

Object D is an extended source with semi-major and semi-minor axes of $A=0.22\arcsec$\ and $B=0.053\arcsec$, which leads to SN offsets of $R=4.9$ and 4.1, in F606W and F814W respectively. As $R<5$, this SN\,Ia cannot be considered truly hostless unless we can show that object D is unlikely to be a cluster member. In the natural system of \textit{HST} ACS, object D is $m_{\rm F606W}=25.56$ and $m_{\rm F814W}=25.75$ mag, and has a color $= -0.20$ mag. As shown in Figure \ref{fig:A2495cmd} it consistent with a red sequence cluster galaxy. We use Sirianni et al. (2005) to convert this photometry into Landolt filter system and find that object D is $m_V \approx 25.6$ and $m_I \approx 24.5$ mag. We apply the distance modulus of Abell 2495 ($\mu \approx 37.8$ mag) and find that intrinsically, object D is $M_V \approx -12.2$ and $M_I \approx -13.3$ mag. This is brighter than the limiting magnitudes quoted for the CFHT deep stack of Abell 2495, but those limits are for point-like sources and object D is extended -- in fact, it is just barely visible as an extended source in Figure 3 of Sand et al. (2011). 

Object D is clumpy and has an ellipticity of 0.7, which is higher than the ellipticity of the bright red sequence galaxies we identify in Abell 2495. Morphologically, object D resembles an inclined disk galaxy -- can we use the disk scale length to assess whether it may be a blue spiral galaxy at higher redshift? Disk galaxies are generally well fit by an exponential function for the flux intensity as a function of radius, $I(R) = I_0 e^{-R/R_d}$, where $R_d$ is the characteristic disk scale length. We estimate $R_d\lesssim0.4\arcsec$, with no attempt to de-project or account for inclination. If object D is a cluster member, this corresponds to $R_d\sim0.6$ kpc, which is roughly appropriate for a disk galaxy. In Section  \ref{ssec:chance} we also discuss the relatively large probability that this is a chance alignment, but formally we cannot exclude the possibility that object D is a cluster member and the host of the SN in Abell 2495.

\subsubsection{Abell 399}\label{sssec:a399}

In Figure \ref{fig:coregA399} we see a large elliptical galaxy near the location of the SN\,Ia. Sand et al. (2011) describes how the SN is $>7 R$ from the large elliptical, and how the SN's and galaxy's line-of-sight velocities differ by $3\sigma$, indicating they are not associated. We use these \textit{HST} images to re-evaluate the parameter $R$ from Equation \ref{eq:r}, which describes the SN's offset in terms of the galaxy's isophotes. We find that $R\sim6$ in both F606W and F814W, re-establishing the SN as independent of the stellar population of this large neighbor galaxy.

\begin{figure}
\begin{centering}
\includegraphics[trim=0.0cm 0.0cm 0.0cm 0.0cm, clip, width=8.5cm]{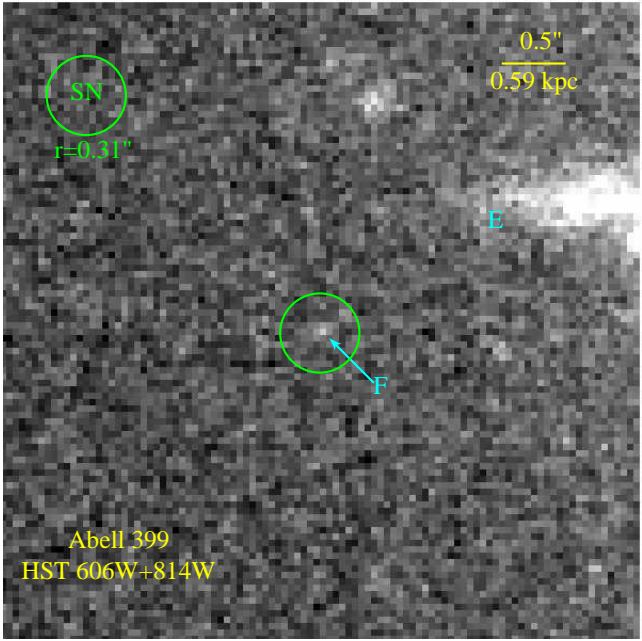}
\caption{The \textit{HST}+ACS image from the deepest stack of both filters F606W and F814W for the candidate intracluster SN\,Ia in Abell 399; a zoom-in of Figure \ref{fig:coregA399}. \label{fig:coregA399zoom}}
\end{centering}
\end{figure}

In Figure \ref{fig:coregA399} we see a small, faint source at the center of the green circle marking the SN\,Ia's position, labeled object F in Figure \ref{fig:coregA399zoom}. Object F is detected with low significance, but with coincidence in both the F606W and F814W images. It is a point-like source, with a FWHM of $\sim3$ pixels, and is offset by $<1$ pixel from the SN\,Ia location. Object F is therefore very likely to be physically associated with the SN\,Ia in Abell 399 -- the small probability of a chance alignment with a foreground star or background host is discussed in Section \ref{ssec:chance}. In Figure \ref{fig:coregA399zoom} we also label the next-nearest object E; it is a clumpy extended source, but we find it far less likely to be physically associated with the SN and so we focus our attention on object F.

Object F is $m_{\rm F606W}=28.7\pm0.3$ and $m_{\rm F814W}=28.7\pm0.2$ mag, below our limiting magnitudes as shown in Figure \ref{fig:de}. Its photometry is consistent with a red sequence galaxy in Abell 399, which is very similar to the red sequence shown for Abell 2495 in Figure \ref{fig:A2495cmd}. We use Sirianni et al. (2005) to convert this photometry into the Landolt filter system and find that object F is $m_V\approx 28.8$ and $m_I\approx 27.4$ mag. We apply the distance modulus of Abell 399 ($\mu\approx 37.2$ mag) and find that intrinsically, object F is $M_{V}\approx -8.4$ and $M_{I}\approx -9.8$ mag, and has a color of $V-I \approx 1.4$. This is also well matched to the expected magnitude and color of a bright GC: the luminosity function for GCs is a Gaussian that peaks at $M_I\approx-7.4$ and has $\sigma\approx1.2$, and GC colors span from $M_V-M_I\approx0.7$--$1.5$ (e.g., West et al. 2011). Although GCs have been found distributed between galaxies in rich clusters (e.g. Peng et al. 2011; West et al. 2011), their density drops with clustercentric radius and we would not expect a significant number of intracluster GCs at the location of this SN\,Ia, 616 kpc from the BCG. However, the presence of the nearby elliptical galaxy (Figure \ref{fig:coregA399}) makes the GC hypothesis more likely.

In Section \ref{ssec:disc.rates}, we show that the SN\,Ia rate enhancements in dwarf galaxies or GC implied by the nature of object F are within all existing theoretical and observational limits. If we assume that the rate per unit mass of SNe\,Ia is not much larger in dwarf galaxies vs. GC (or vice versa)  -- which would put a strong prior on the nature of object F -- we can estimate the expected surface densities of dwarf red sequence galaxies and red GCs at this location in order to assess which is more probable. Based on our extrapolation of the cluster luminosity function for Abell 399 (Section \ref{ssec:fL}), we estimate there to be $\sim1.5\times10^9$ $\rm L_{\odot}$ in dwarf cluster galaxies at $M_R > -10$ mag. That is about 100--500 dwarf galaxies of 5--10 $\times10^6$ $\rm L_{\odot}$ within a clustercentric radius of $\sim1$ Mpc, or 1--5 $\times 10^{-4}$ dwarfs $\rm kpc^{-2}$. This would apply at any location in the cluster, and so applies for the location of object F. To assess whether satellites of the nearby elliptical galaxy might raise the predicted surface density at this location, we use the radial distribution of low-redshift, low-mass satellite galaxies presented by Prescott et al. (2011; see their Figure 5). The location of the SN in Abell 399 is $R\approx10$ $\rm kpc$ from the center of the nearby elliptical; at this radius Prescott et al. (2011) find that the surface density of satellites is $\sim7\times10^{-6}$ $\rm kpc^{-2}$. The caveat here is that they consider only isolated primary galaxies, and so their results represent an upper limit on the radial distribution we could expect in rich clusters. However, as this is much lower than what we expect from the cluster luminosity function we conclude that object F is unlikely to be a satellite galaxy of the nearby elliptical.

We can estimate the surface density of GC from the nearby elliptical galaxy (Figure \ref{fig:coregA399}), which has absolute magnitudes of $M_V\approx-19.5$ and $M_I\approx-20.7$, and a stellar mass of $\sim7.6\times10^9$ $\rm M_{\odot}$ (Bell et al. 2003). A galaxy of this stellar mass has between 10--100 GC within $R<50$ $\rm kpc$, and the surface density radial distribution for GC is $N(r) = N_0 r^{-2.4}$ $\rm GC\ kpc^{-2}$ (Zaritsky et al. 2015); $\sim75\%$ of the GC are internal to 10 $\rm kpc$ (the distance of the SN in Abell 399 from the nearby elliptical). Normalizing the radial distribution to a total of 10--100 GC gives $N\approx 0.3-3\times10^{-2}$ $\rm GC\ kpc^{-2}$. Based on the GC luminosity function of West et al. (2011) $\sim84\%$ of all GC are below our limiting magnitudes, and so the {\it observable} surface density at 10 $\rm kpc$ is lowered to $0.5-5\times10^{-3}$ $\rm GC\ kpc^{-2}$. This is higher than our estimated surface density for cluster dwarf galaxies, suggesting it is more likely to see a GC at the location of object F than a cluster dwarf. However, as our lower limit for the GC surface density is equal to our upper limit on the dwarf galaxy surface density the estimates are not significantly different enough to make a robust claim to the nature of object F. The caveat here is that we have used a radial distribution for {\it all} GC, but the population of red GCs has a significantly shorter radial extent than blue GCs -- in fact, for cluster ellipticals the radial distribution of red GCs appears to be truncated near the effective radius of the parent galaxy (e.g., Brodie \& Strader 2006).

In summary, we find that the SN\,Ia in Abell 399 is likely physically associated with object F. Without any constraints from expected SN\,Ia rates, we find that object F is less likely to be a dwarf galaxy than a GC from the nearby elliptical. We discuss the probability of chance alignments in Section \ref{ssec:chance}, and the possibility that object F is lingering emission from the SN\,Ia in Section \ref{ssec:snlim}. We discuss the implications of object F for SN\,Ia rates in dwarf galaxies and GCs, and whether these implications provide a prior on the nature of object F, in Section \ref{ssec:disc.rates} and the impact of object F on using SN\,Ia as tracers of the ICL at high-redshift in Section \ref{ssec:disc.ICL}.

\subsubsection{Abell 85}\label{sssec:a85}

In Figure \ref{fig:coregA85}, we see that the location of the SN\,Ia appears devoid of objects. For the nearest source, labeled object G in Figure \ref{fig:coregA85}, the SN is $R\sim13$ away. We ran Source Extractor with very relaxed parameters, but found that the few sources detected are consistent with noise peaks (i.e., they had very low significance, their detections in F814W only and were not coincident with any F606W low-significance detections, and were not visually confirmed in the F606W+F814W stack). We conclude that the SN\,Ia in Abell 85 is truly hostless.

\subsection{Probabilities of Random Line-of-Sight Alignments}\label{ssec:chance}

In our discussion of the nature of object F in Abell 399 we estimated the surface densities of faint cluster objects such as dwarf red sequence galaxies and GCs to be $\sim5\times10^{-4}$ objects per $\rm kpc^{2}$. The probability of a faint cluster object appearing randomly within the positional uncertainty of our IC SNe\,Ia is negligible, $\lesssim0.0003\%$. The presence of object F is not a chance alignment with a cluster object.

We used TRILEGAL (Girardi et al. 2005) to simulate a foreground star population in the directions of our four fields to a limiting magnitude of $m_{\rm V}=29$. The probability of a star appearing randomly within 0.2\arcsec\ of our IC SN Ia coordinates is $\leq$0.02\%. We judge that it is extremely unlikely that object F at the location of the SN in Abell 399 is a foreground red star.

We used the Hubble Deep Field catalogs (Williams et al. 1996) to simulate a population of faint objects between $25.0 < m_{\rm F814W} < 29.5$. This is the magnitude range in which we detect sources in our {\it HST} images but not our CFHT images. The new sources detected near the SN\,Ia locations in Abell 2495 and Abell 399 fall in this magnitude range. We find the probability of a faint field object randomly being within 0.31\arcsec\ and 0.34\arcsec, the positional uncertainties on the locations of the SNe\,Ia, are $\sim2.0\%$ and $\sim2.5\%$, respectively. In our original work $\sim2\%$ of the cluster stellar mass in faint dwarf galaxies was below our detection limit (Sand et al. 2011). Assuming the SN\,Ia occurrence rate per unit mass is equivalent in all populations (high- and low-mass galaxies, and intracluster stars), this means $\sim2\%$ of the all MENeaCS SNe\,Ia should be hosted by undetected, faint dwarf galaxies ($\sim0.5$ out of $23$ SNe\,Ia). We now see that this is approximately the same chance as finding a background galaxy within the positional uncertainty. However, object F in Abell 399 is not just within the positional uncertainty, but appears within 0.05\arcsec\ (1 ACS pixel) of the SN location. The probability of random alignment with a background object within 0.10\arcsec\ is just 0.3\%. On the other hand, if we increase the radius to 1.0\arcsec\ (i.e. the distance encompassing nearby objects for Abell 2495), the probability of chance alignment increases to $\sim18\%$. 

As a final, alternative estimate we use our own source catalogs for all four $HST$-ACS fields and find that the fraction of our imaged area within $R<5$ of an object is $\sim3\%$. While it remains very unlikely that object F is a chance alignment with the SN\,Ia in Abell 399 (Figure \ref{fig:coregA399zoom}), we cannot say the same for the objects near the SN\,Ia location in Abell 2495 (Figure \ref{fig:coregA2495zoom}).

\subsection{Limits on the SNe\,Ia and/or its Companion}\label{ssec:snlim}

A normal SN\,Ia has faded $\sim4$ magnitudes by $\sim100$ days after peak brightness. After this time the decline is set by the decay rate of Co$^{56}$, and the SN\,Ia continues to fade at $\sim 1$ magnitude per 100 days in $I$-band, and $\sim 1.3-1.5$ mag per 100 days in $BVR$ (e.g. as seen for normal twin SNe\,Ia 2011fe and 2011by; Graham et al. 2015a). For a normal SN\,Ia such as SN\,2011fe, the intrinsic brightness at $\sim1000$ days is $V\sim-5$ magnitudes (e.g., Kerzendorf et al. 2014; Graham et al. 2015b). Most of the flux comes from blue emission features of [\ion{Fe}{2}] at $<6000$ \AA\ (e.g., Taubenberger et al. 2015; Graham et al. 2015b), and the late-time $V-I$ color of a normal SN\,Ia is expected to be $\sim0$. After $\sim1000$ days, the predicted rate of decline for normal SNe\,Ia is even slower (Seitenzahl et al. 2009), and so $V\sim-5$ mag is a conservative upper limit on SN\,Ia brightness after 1000 days.

Over-luminous SNe\,Ia that resemble SN\,1991T have been observed to decline more slowly. For example, SN\,1991T itself was $V\sim-10$ at $\sim600$ days (Cappellaro et al. 1997) and SN\,2000cx had a $V$-band slope of $\sim0.65$ mag per 100 days at $\sim700$ days after peak brightness (Sollerman et al. 2004), so a SN\,1991T-like event could be $V\lesssim-7$ at 1000 days. However, none of the MENeaCS SNe\,Ia were spectroscopically similar to SN\,1991T, and are furthermore unlikely to belong to this subclass because 91T-like SNe\,Ia are associated with younger stellar populations (Howell et al. 2009). Although less is known about the late-time decline of sub-luminous SNe\,Ia, SN\,1991bg itself was already $V\approx-6$ by $\sim600$ days after peak brightness (Turatto et al. 1996). Our intracluster SN\,Ia in Abell 85 was classified as SN\,1991bg-like, but no object is detected at its location in the {\it HST} ACS imaging.

Could we detect the emission from a shocked companion star? For normal SNe\,Ia, theoretical predictions for a non-degenerate companion shocked by the SN\,Ia ejecta include an increase in temperature and luminosity, up to $10^3$ to $10^4$ $\rm L_{\odot}$ by 1--10 years after explosion (Pan et al. 2013; Shappee et al. 2013). Such a companion would be blue, and have $V>-4$ mag, which is well below our limiting magnitudes. A light echo is also likely to be blue, similar to the color of a SN\,Ia at peak light, and even fainter. In order to formally rule out the possibility that the objects identified near the SN locations in Abell 2495 and Abell 399 are lingering emission from the SN\,Ia or its binary companion, we discuss each in turn. 

\textbf{Abell 2495 --} In Figure \ref{fig:coregA2495zoom} we identify object C as a possible point source near the location of the SN\,Ia in Abell 2495. The spectrum of this SN\,Ia was obtained on 2009-06-18.58 UT at Gemini Observatory as part of the MENeaCS follow-up campaign, and we classified it as a normal SN\,Ia at $\sim+3$ months after peak brightness. At the time of our {\it HST} ACS images obtained on 2013-10-04, the SN\,Ia would be +1569 days old (4.3 years). At this time, object C is $m_{\rm F606W} = 29.8 \pm 0.2$ and $m_{\rm F814W} = 29.0 \pm 0.2$ mag, or $m_V\approx 30.0$ and $m_I\approx27.8$ mag, in our {\it HST} ACS imaging. We apply the distance modulus of Abell 2495 ($\mu\approx 37.8$ mag) and find that intrinsically, object C is $M_{V}\approx -7.8$ and $M_{I}\approx -10.0$ magnitudes. This is both significantly brighter and redder than a normal SN\,Ia is predicted to be at $\sim4$ years. We conclude that object C is unlikely to be the SN\,Ia or an evolved companion star.

\textbf{Abell 399 --} In Figure \ref{fig:coregA399zoom} we identify object F as a point source at the location of the SN\,Ia in Abell 399. The classification spectrum of this SN\,Ia, obtained on 2008-11-28.49 UT at Gemini Observatory as part of the MENeaCS follow-up campaign, showed it to be a normal SN\,Ia at $\sim+2$ weeks after peak brightness. In our {\it HST} ACS images obtained on 2013-12-04, the SN\,Ia would be +1832 days old (5 years). This object is $M_{V}\approx -8.4$ and $M_{I}\approx -9.8$ mag, has a color of $V-I \approx 1.4$, both significantly brighter and redder than a normal SN\,Ia is predicted to be at extremely late times. We conclude that object F is unlikely to be the SN\,Ia or an evolved companion star.

\section{Discussion}\label{sec:disc}

Our analysis of the {\it HST} ACS images at the locations of our 4 intracluster MENeaCS SNe\,Ia has shown that one is hosted by either a dwarf red sequence galaxy or red GC (Abell 399); one is potentially associated with a nearby spiral or irregular galaxy consistent with the red sequence but also has a relatively high probability of being a chance alignment (Abell 2495); and two appear to be truly hostless (Abell 1650 and Abell 85).  We discuss the implications of our results for the rates of SNe\,Ia in faint cluster hosts in Section \ref{ssec:disc.rates} and for the use of SNe\,Ia as tracers of the ICL in Section \ref{ssec:disc.ICL}.

\subsection{Implications for SN\,Ia Rates in Clusters}\label{ssec:disc.rates}

In Section \ref{sec:ana} we found that the SN\,Ia in Abell 399 was likely hosted by the faint object F, and that this source is consistent with being either a cluster dwarf galaxy or a GC. Here we consider the implications of both scenarios on the rate of SNe\,Ia in dwarf galaxies and GCs, and whether established SN\,Ia rates (or limits) in these populations can constrain the physical nature of object F. 

\subsubsection{Dwarf Galaxies}

If object F in Abell 399 is a dwarf galaxy, does this imply a significantly enhanced SN\,Ia rate in faint cluster galaxies? The rate per unit mass in a population, $\mathcal{R}$, is expressed by $\mathcal{R} = C \times N / M$, where $C$ is a detection efficiency, $N$ is the number of SNe\,Ia, and $M$ is the mass in the population. As described in Sand et al. (2011), our original CFHT deep stacks left $\lesssim2\%$ of the mass in faint cluster galaxies undetected, but we now believe that population has hosted 1 SN\,Ia. We can estimate the implied \textit{relative} rate per unit mass in the faintest $2\%$ of cluster galaxies with the following equation:

\begin{equation}\label{eq:rratio}
\frac{\mathcal{R}_{\rm 2\%}}{\mathcal{R}} = \frac{C_{\rm 2\%} \times N_{\rm 2\%} / M_{\rm 2\%}}{C \times N / M}.
\end{equation}

\noindent
Sand et al. (2011) describes how a small difference in MENeaCS detection efficiencies between hosted and hostless SNe\,Ia are introduced by two effects: (1) it is more difficult to detect transients on top of a host galaxy (even with difference imaging techniques), and (2) the spectroscopic follow-up coverage for the hostless population was slightly more extensive than for the hosted SNe (they were run under separate proposals). Together, this difference works out to be $C_{\rm 2\%} = 1.2 C$. Sand et al. (2012) present that the number of SNe\,Ia hosted by all cluster galaxies within 1 Mpc is $N=11$, and so with $N_{\rm 2\%}=1$ we find that $\mathcal{R}_{\rm 2\%}/\mathcal{R} \approx 5.5$. Repeating this calculation using only red sequence cluster galaxies yields a similar rate enhancement because the number of SNe\,Ia in red sequence members within 1 Mpc is $N_{\rm RS}=6$ (i.e., $0.5 N$), the stellar mass in red sequence galaxies is $M_{\rm RS} \sim 0.5 M$, and $C_{\rm RS}=C$.

Ultimately this potential rate enhancement by a factor of $\sim5$ is quite uncertain, as it is based on just one SN\,Ia and an indirect estimate of the amount of mass in faint cluster galaxies. As introduced in \S~\ref{ssec:intro.baryons} and \ref{ssec:chance}, by assuming the SN\,Ia rate per stellar mass is equal in all cluster populations we estimated that the expectation value for the number of MENeaCS SNe\,Ia in dwarf hosts is $\sim0.5$. For a more restrictive estimate of the expectation value, we limit to red sequence galaxies within the clustercentric radius of 1 Mpc used above. Assuming the rate in bright cluster red sequence galaxies ($\sim50\%$ of the stellar mass hosting $N_{\rm RS}=6$ SNe\,Ia) is the same as that in dwarf red sequence galaxies ($\sim2\%$ of the stellar mass), the expectation value is $\sim0.24$ SNe\,Ia in dwarf hosts. With Poisson statistics the probability of observing $\geq1$ SN\,Ia given this expectation value is 0.16, and the $1\sigma$ uncertainties our estimated rate enhancement are $5.5^{12.7}_{-4.5}$ (Gehrels 1986) -- consistent with {\it no} rate enhancement. Furthermore, if the luminosity function is steeper than expected there could be more than 2\% of the stellar mass residing in such faint red galaxies. We plan to use our deep \textit{HST} images to constrain the faint-end slope of the cluster luminosity function in later work, but consider it beyond the scope of this paper.

The potential SN\,Ia rate enhancement by a factor of $\sim5$ in the faintest dwarf galaxies is not too large to be unphysical, as the SN\,Ia rate is known to vary by factors of that size (and more) with host galaxy properties such as the specific star formation rate (e.g. Mannucci et al. 2005; Scannapieco \& Bildsten 2005; Sullivan et al. 2006; Smith et al. 2012). This potential rate enhancement is also not so large that we expect to have already observed it in wide-field sky surveys such as SDSS. For example, Smith et al. (2012) find that the SN\,Ia rate per unit mass decreases as a function of host stellar mass, but their lowest stellar mass bin is $\sim5\times10^8$ $\rm M_{\odot}$ and the uncertainty on its rate is a factor of $\sim2$--3. Furthermore, they mention that $\sim2\%$ of the SNe\,Ia in their sample have undetected host galaxies. Current and future wide-field surveys such as the Palomar Transient Factory and the LSST should be able to improve the SN\,Ia rate in faint dwarf galaxies (e.g., Conroy \& Bullock 2015).

\subsubsection{Globular Clusters}

If object F in Abell 399 is a GC, it would be the first confirmed GC to host a SN\,Ia. As GCs are purely old stellar populations ($>2$ Gyr), such an association would also be direct confirmation that SNe\,Ia progenitors include truly old star systems. The fraction of galaxy stellar mass in GC systems is $\sim 2\times10^{-3}$ for elliptical galaxies (Harris et al. 2009; Peng et al. 2008; Zaritsky et al. 2015). Based on this and the total stellar mass in red sequence galaxies in MENeaCS clusters, we estimate that $\gtrsim 10^{12} M_{\odot}$ in GCs has been surveyed by MENeaCS. If object F is a GC and has produced a SN\,Ia, it implies a rate $\sim$25 times higher than the rate in our cluster red sequence galaxies from Sand et al. (2012). This is on a similar scale to theoretically predicted enhancements due to dynamical interactions in the dense stellar environments of GCs (Pfahl et al. 2009), and below the current limits placed by non-detections of GC at the locations of SNe\,Ia in archival HST images ($\lesssim42\times$; Voss \& Nelemans 2012; Washabaugh \& Bregman 2013).

Would such an enhanced rate in GCs have already been noticed? Potentially not. It would imply that $\lesssim5\%$ of the SNe\,Ia in ellipticals are hosted by their GCs, but as we discussed in Section \ref{sssec:a399} the radial distribution of GCs follows the galaxy light profile and drops off at $\sim5 R$. It is entirely conceivable that GC-hosted SNe have simply been assigned as belonging to the parent galaxy. This might lead to a relative rate enhancement in the outer regions of ellipticals, which could be measured and attributed to GCs. However, such an effect would be difficult to confidently measure for two reasons. First, there is a detection bias of SNe being easier to find when they are not embedded deep in the host galaxy. Second, a stellar population originating in GCs -- able to produce an enhanced rate of SN\,Ia -- may have previously released into the halos (or bulges) of galaxies from GCs due to collisions with clouds or other GCs, winds from supernovae in the GC, tidal forces for GC on elliptical orbits about their host, evaporation of stars during interal GC relaxation, and/or dynamical friction (Fall \& Rees, 1985). It is conceivable that these issues combined could conspire to blur the signal of a SN\,Ia rate enhancement in GCs in the radial distribution of SNe\,Ia.

\subsubsection{Rates Summary}

We find that the implied SN\,Ia rate enhancements in either dwarf galaxies or GCs do not conflict with existing observations or theoretical predictions, but also do not provide a means to constrain the nature of object F. Additionally, if either of the 2 SNe\,Ia in Abell 1650 and Abell 85 -- which appear hostless in our deep \textit{HST} images -- were in fact hosted by the $\sim0.2\%$ of the stellar mass in small galaxies that remains below our limiting magnitudes, then by Equation \ref{eq:rratio} this indicates a rate enhancement by a factor of $\sim55$ in the faintest dwarf galaxies. This is conspicuously high and would have been noticed in wide-field surveys. We therefore conclude that these 2 SNe\,Ia were truly hosted by the population of intracluster stars stripped from their host galaxy and residing in the cluster's gravitational potential.

\subsection{Implications for IC SNe\,Ia as Tracers of the ICL}\label{ssec:disc.ICL}

Sand et al. (2011) reported that of the 23 cluster SNe\,Ia discovered by MENeaCS, 4 had no apparent host galaxy in deep CFHT images that left just $\sim2\%$ of the total cluster stellar mass in undetected faint galaxies. With our deep $HST$ imaging we find that 1 out of the 23 cluster SNe\,Ia, $\sim4\%$, is hosted by a faint point-source (object F in Abell 399). Finding $\sim4\%$ of the SNe\,Ia hosted by $\sim2\%$ of the stellar mass is not surprising, but here we take a closer look at how $f_{\rm ICL}$ is derived in order to confirm the utility of SNe\,Ia as tracers of the ICL at higher redshift.

The fraction of intracluster light, $f_{\rm ICL}$, is calculated by dividing the number of hostless SNe\,Ia by the total number of cluster SNe\,Ia (hosted+hostless) discovered by MENeaCS (Sand et al. 2011). To do this, the detection efficiencies must be equal for hosted and hostless SNe\,Ia; in other words, a survey must apply the same discovery and spectroscopic classification constraints to both populations, or be able to account for any bias in the pipeline. This was true for all MENeaCS SNe\,Ia except the IC SN in Abell 2495, which was preferentially observed with Gemini spectroscopy despite being fainter than the magnitude limit applied to follow-up of MENeaCS SNe. For this reason, the hostless SN in Abell 2495 was not included in the calculation of the fraction of IC stellar mass in Sand et al. (2011). In order to combine only the same physical regions of each cluster, they limit to a radius $<R_{\rm 200}$ (i.e., the virial radius); all four apparently hostless SNe\,Ia are within $R_{\rm 200}$. Using the remaining 3 apparently hostless SNe\,Ia as intracluster, and the 13 hosted SNe\,Ia within $R_{\rm 200}$, Sand et al. (2011) measure $f_{\rm ICL} = 0.16^{+0.13}_{-0.09}$.

Given the proximity of the SN\,Ia in Abell 399 to a large nearby galaxy (see Section \ref{sssec:a399}), they also repeat this calculation assuming that this SN\,Ia was hosted, and find $f_{\rm ICL} = 0.11^{+0.12}_{-0.07}$. In light of the faint object F at the location of the SN\,Ia in Abell 399, we can now say the latter is the more accurate measurement of $f_{\rm ICL}$. Although this distinction may not seem significant because the difference between these two $f_{\rm ICL}$ measurements is within their relatively large statistical errors, future facilities such as the Large Synoptic Survey Telescope (LSST) will generate bigger sample sizes and have smaller uncertainties and a better understanding of the fraction of apparently hostless SNe\,Ia will be needed in this regime. LSST itself will be able to provide this because its deep stack images have a projected detection limit of $r\sim27.5$ mag\footnote{http://lsst.org/lsst/science/science\_portfolio}.

Until then, we can only caution that future surveys will have to consider that 25--30\% of apparently hostless SNe\,Ia might not belong to the IC stellar population. Is that too large for hostless SNe\,Ia to be scientifically useful tracers of $f_{\rm ICL}$ at higher redshifts? Some numerical simulations indicate that $f_{\rm ICL}$ grows with cosmological time, and by $z\sim0$ is $\sim2\times$ larger than at $z\sim0.4$ (e.g. Murante et al. 2007); others find that most IC stars are stripped at $z > 1$, in which case $f_{\rm ICL}$ would remain constant since then (Puchwein et al. 2010). Direct measurements of this low surface brightness component have shown that $f_{\rm ICL}\lesssim25\%$ at $z\lesssim0.1$ and $\sim10\%$ at $z\sim0.2$ (e.g. Gonzalez et al. 2007; Zibetti et al. 2005), but HST imaging of $0.4 < z < 0.8$ clusters has found no evolution in $f_{\rm ICL}$ since $z < 0.8$ (Guennou et al. 2012). We therefore surmise that assessments of $f_{\rm ICL}$ from SNe\,Ia will require uncertainties of $<30\%$ in order to compare with some theoretical models and the direct surface brightness measurements. This sounds discouraging, but there is hope: below, we suggest that the best way to improve this uncertainty and use hostless SNe\,Ia as high-redshift ICL tracers is to constrain the SN\,Ia occurrence rate in faint hosts. 

Dwarf galaxies represent a small fraction of the total cluster stellar mass, which is constrained by measurements of the galaxy luminosity function. If their SN\,Ia occurrence rate is equal to that in elliptical galaxies, then measurements of $f_{\rm ICL}$ can account for the contamination from apparently hostless SNe\,Ia in dwarf hosts. However, in the preceding section we show that if object F is a cluster dwarf, the SN\,Ia occurrence rate in the faintest $\sim2\%$ of cluster galaxies could be up to $\sim5\times$ higher than in elliptical galaxies. If confirmed, the number of SNe\,Ia in faint dwarfs could be up to half of all apparently hostless SNe\,Ia.

Compared to dwarf galaxies, GC represent an even smaller fraction ($\lesssim 0.002$) of the total stellar mass in clusters -- much less than the fraction of intracluster stars ($\sim 0.16$). If their SN\,Ia occurrence rate is equal to that in elliptical galaxies, accidentally including the very few SNe\,Ia in GC at large radial offsets from their host as part of the ICL will have a negligible effect on measurements of $f_{ICL}$. (Most SNe\,Ia in GC will be associated with elliptical galaxies, because GC have a radial distribution similar to that of stars.) On the other hand, if the SN\,Ia occurrence rate in GC is $25\times$ that in elliptical galaxies -- as implied if object F is a GC -- then up to $\sim5\%$ of all cluster SNe\,Ia, and up to $\sim30\%$ of the apparently hostless cluster SNe\,Ia, may actually be associated with GCs.

In this work we have shown that $\sim75\%$ of all apparently hostless SNe\,Ia are truly intracluster, and that contamination is only a significant problem if the SN\,Ia occurrence rate in low-mass galaxies or GC is {\it enhanced}. This issue can be resolved by SN\,Ia rates analyses from low-redshift, wide-field surveys such as the Palomar Transient Factory, SDSS, or LSST, combined with deep imaging for a larger sample of low-redshift apparently hostless SN\,Ia. Such an effort should sufficiently constrain the SN\,Ia rate in dwarfs and GCs such that extremely deep imaging will not be required to confirm all apparently hostless cluster SNe\,Ia, and facilitate their use as tracers of the ICL to higher redshifts.

\section{Conclusion}\label{sec:conc}

We have presented deep \textit{HST}+ACS images at the locations of 4 IC SNe\,Ia in rich galaxy clusters, obtained $>$3 years after explosion. This is the largest single-survey sample of IC SNe\,Ia in rich clusters, and these data are the deepest images ever obtained at the locations of IC SNe\,Ia, lowering the amount of stellar mass left undetected in cluster galaxies from $\sim2\%$ to just $\sim0.2\%$ ($\lesssim0.005\%$ if we assume a shallow faint-end slope for the galaxy luminosity function,  $\alpha_d=-1.0$). We have confirmed that the 2 SNe\,Ia in Abell 1650 and Abell 85 are hostless, and truly belong to the intracluster stellar population of stars stripped from their host galaxy and residing in the cluster potential. This indicates that at least some SN\,Ia progenitors have truly old progenitor stars ($>2$ Gyr). We could not rule out that the SN\,Ia in Abell 2495 was hosted by a nearby disk galaxy that has a magnitude, color, and size consistent with cluster membership, but also found a relatively high probability that this is a random association. We have shown that the SN\,Ia in Abell 399 was very likely hosted by a faint red point-like source that has a magnitude and color consistent with both dwarf red sequence galaxies and red GCs. Our statistical analysis of the expected surface densities has shown that a dwarf galaxy is less likely at that location than a GC, due to the presence of a nearby elliptical galaxy. We have demonstrated that the rate enhancements in dwarfs or GCs implied by this new faint host are plausible under current observational constraints, and we do not reject either hypothesis. We have also explicitly ruled out the possibility that we could observe extremely late-time emission from the SNe\,Ia themselves or a possible shocked companion.

Our discovery of a faint host galaxy for 1 of the 4 SNe\,Ia that appeared to be hostless is a potential problem for measuring the evolution in the fraction of intracluster light across cosmic time, because $f_{\rm ICL}$ will need to have uncertainties $<25\%$ in order to distinguish between models or compare with independent measurements. For this reason we argue that hostless SNe\,Ia in rich clusters should be used to measure $f_{\rm ICL}$ with caution, and that it would be best if deep imaging for a larger sample of low-redshift, apparently hostless SNe\,Ia were obtained in order to better constrain the rate in dwarf galaxies and GCs.

\acknowledgments

This work is based on observations made with the NASA/ESA Hubble Space Telescope and obtained from the Mikulski Archive for Space Telescopes at the Space Telescope Science Institute, which is operated by the Association of Universities for Research in Astronomy, Inc., under NASA contract NAS 5-26555. These observations are associated with program \#12937 and partially funded by a NASA grant. We thank the $HST$ staff for their assistance with this program. 

This work is also based in part on observations obtained with MegaPrime/MegaCam, a joint project of CFHT and CEA/DAPNIA, at the CanadaÐFranceÐHawaii Telescope (CFHT) which is operated by the National Research Council (NRC) of Canada, the Institut National des Science de lÕUnivers of the Centre National de la Recherche Scientifique (CNRS) of France, and the University of Hawaii. We remain grateful to the operators of the CFHT queue without whom the original MENeaCS project would not have been possible.

MLG's position in the supernova research group at U.C. Berkeley is supported by Gary \& Cynthia Bengier. MLG and DJS were supported during \textit{HST} proposal preparation by the Las Cumbres Observatory Global Telescope Network. CJP is supported by the Natural Sciences and Engineering Research Council of Canada. We thank our anonymous reviewer for their constructive comments and useful suggestions.

\bibliographystyle{apj}
\bibliography{apj-jour,mybib}

\end{document}